\documentstyle[pre,twocolumn,aps,psfig]{revtex}
\begin{document}
\preprint{}
\draft
\twocolumn[\hsize\textwidth\columnwidth\hsize\csname@twocolumnfalse%
\endcsname
\title{Discrimination of the Healthy and Sick Cardiac Autonomic
Nervous System by a New Wavelet Analysis of Heartbeat Intervals}
\author{Y. Ashkenazy$^{\text{a,b}}$, M. Lewkowicz$^{\text{c,a}}$,
J. Levitan$^{\text{c,a}}$,\\
H. Moelgaard$^{\text{d}}$, P.E. Bloch Thomsen$^{\text{e}}$ and
K. Saermark$^{\text{f}}$}
\address{
(a) Dept. of Physics, Bar-Ilan University, Ramat-Gan, Israel\\
(b) Gonda Goldschmied Center, Bar-Ilan University, Ramat-Gan, Israel\\
(c) College of Judea and Samaria, Ariel, Israel\\
(d) Dept. of Cardiology, Skejby Sygehus, Aarhus University Hospital, Denmark\\
(e) Dept. of Cardiology, Gentofte Amtsygehus, Copenhagen University Hospital,
Denmark\\
(f) Dept. of Physics, The Technical University of Denmark, Lyngby, Denmark.}
\date{\today}
\maketitle
\begin{abstract}
{
We demonstrate that it is possible to distinguish with a complete
certainty  between healthy subjects and patients with
various dysfunctions of the cardiac nervous system by way of
multiresolutional wavelet transform of RR intervals. We repeated the study of
Thurner {\it et al} on different ensemble of subjects. We show that
reconstructed series using a filter which discards wavelet coefficients
related with higher scales enables one to classify individuals for
which the method otherwise is inconclusive. We suggest a delimiting
diagnostic value of the standard deviation of the filtered, reconstructed RR
interval time series in the range of $\sim  0.035$ (for the above
mentioned filter), below which individuals are at risk.
}
\end{abstract}
\pacs{
}
]
\narrowtext
\newpage
\section{Introduction}
Measurement of heart rate (HR) and evaluation of its rhythmicity have been used
for a long time as a simple clinical indicator \cite{Moelgaard95}.
The main adaptive regulation of the sinus node function and thereby the HR,
is exerted by the autonomic nervous system. The sinus node of the heart is a
major organ in the integrated control of cardiovascular function. HR
abnormality may therefore be an early or principle sign of disease or
malfunction.

Research from the last decade indicates that a quantification of the discrete
beat to beat variations in HR - heart rate variability (HRV) may be used more
directly to estimate efferent autonomic activity to the heart and the
integrity of this cardiovascular control system \cite{Furlan90}. The finding
that power
spectral analysis of HRV could be used as a marker of cardiac autonomic
outflow to the heart, was considered a breakthrough for clinical research
\cite{{Akselrod81},{Pomeranz85}}.

Autonomic dysfunction is an important factor in a number of conditions. In
diabetes, an abnormality in autonomic nervous function signals an adverse
prognosis and risk of subsequent heart disease. Recognition of early
dysfunction is therefore important. In overt heart disease autonomic imbalance
 is of significant importance in the pathophysiology of sudden cardiac death.
Abnormal autonomic balance is an important prognostic factor. In heart failure
this control system may be significantly deranged.

Techniques which can discriminate the healthy HRV profile from a sick one are
therefore highly desirable. So far this has not been accomplished, as a
considerable overlap between healthy and sick, (i.e. healthy and diabetes)
\cite{Moelgaard94} or
high and low risk heart disease patients \cite{Bigger91}, have been reported.
The time series
used for HRV analysis are derived from 24-hour ECG recordings. These are
clinically widely used and offer important additional information. However,
several problems have limited the use and interpretation of the spectral
analysis results. The ambulatory time segments inherently lack stationarity.
Furthermore, they often include transients caused by artifacts, ectopic beats,
 noise, tape speed errors which may have significant impact on the power
spectrum \cite{Task96}.
This significantly limits the sensitivity of this technique, and thus may
limit its applicability.

\section{Methods}
One of the most successful techniques to analyze non stationary time series is
the Multiresolution Wavelet Analysis
\cite{{Daubechies92},{Strang96},{NR},{Aldoubri96},{Akay97},{Ivanov96},{Thurner98}}.
This technique was recently utilized in order to analyze a sequence of RR
intervals \cite{{Ivanov96},{Thurner98}}. Ref. \cite{Ivanov96} identifies
different scaling properties in healthy and sleep apnea patients.
In a previous study, Peng {\it et al} \cite{Peng95} were able to distinguish 
between healthy subjects and patients with heart failure by the use of the 
detrended flactuation analysis. Later, 
Thurner {\it et al}
\cite{Thurner98} used a similar procedure but focused on the
values of the variance rather than on the scaling exponent. 
For the scale windows of $m=4$ and $m=5$
heartbeats, the standard
deviations of the wavelet coefficients for normal individuals and heart failure
patients were divided into two disjoint sets.  In this way the authors of ref.
\cite{Thurner98} succeeded to
classify subjects from a test group as either belonging to the heart failure or
the normal group, and that with a 100\% accuracy.

The Discrete Wavelet transform is a mathematical recipe acting on a data vector
of length $2^m$, $m=1,2,\ldots$ and transforming it into a different vector of
the same length. It is based on recursive sums and differences of the vector
components; the sums can be compared with the low frequency amplitudes in the
Fourier transform, and the differences with the high frequency amplitudes. It
is similar to the Fourier transform in respect of
orthogonality and invertibility. The wavelets are the unit vectors i.e., they
correspond to the sine and cosine basis functions of the Fourier transform.
One of the basic advantages of wavelets  is that an event can be simultaneously
described in the frequency domain as well as in the time domain, unlike the
usual Fourier transform where an event is accurately described either in the
frequency or in the time domain. This difference allows a multi resolution
analysis of data with different behaviour on different scales. This dual
localization  renders  functions with intrinsic inaccuracies into reliable data
when they are transformed into the wavelet domain. Large classes of biological
data (such as ECG series and RR intervals) may be analysed by this method.

Heart failure patients generally have very low HRV values. To
further explore the potential possibilities of the Multiresolutional Wavelet
Analysis we have investigated a test group of 33 persons, 12 patients and 21
healthy subjects. The patient group consisted of 10 diabetic patients which
are otherwise healthy and without symptoms or signs of heart disease, one
patient which have had a myocardial infarction and one heart transplanted
patient in whom the autonomic nerves to the heart have been cut.
\begin{figure}[thb]
\psfig{figure=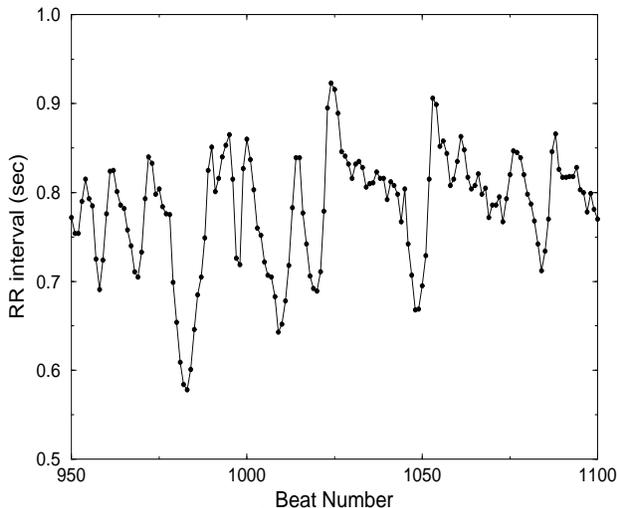,height=7cm,width=10cm,angle=-90}
\caption[]{\label{fig1}
RR interval vs. (heart)beat number for a healthy subject.
}
\end{figure}

We have in the present study applied the same technique as used in ref.
\cite{Thurner98} and have by
Multiresolution Wavelet Analysis been able to identify correctly all but one
of 33 test persons as belonging to the group of healthy subjects or
subjects suffering from  myocardial infarction. The heart
transplanted patient was included as a subject displaying the ultimative
cardiac autonomic dysfunction - complete denervation.

We have, however, elaborated on the procedure applied in ref. \cite{Thurner98}
by utilizing a filter-technique. Thus we perform an Inverse Wavelet Transform,
but retain only a specific scale in the reconstruction of the time series;
a complete separation is observed for $m=4$ or $m=5$. In this way a
reconstructed and filtered time series is obtained and a comparison with
the original time series shows a substantial difference in amplitude between
sick/healthy subjects relative to the difference found in the original
RR interval time series. The choice of $m=4$ or $m=5$ was motivated by the
findings in ref. \cite{Thurner98} and by our own results.

\section{Results}
We have calculated the standard deviation $\sigma_{wave}$ for Daubechies
10-tap wavelet
versus the scale $m$, $1\le m\le 10$, for 33 persons. In accordance with ref.
\cite{Thurner98} we find that for $4\le m\le6$ the $\sigma_{wave}$ separate
the two classes of subjects and hence
provide a clinically significant measure of the presence of cardiac autonomic
dysfunction with
a 97\% sensitivity. This supports in a convincing way the findings of ref.
\cite{Thurner98}. We have been able to confirm this trend with other wavelets.

The main result of this study is however the possibility to display the
standard deviation of the RR interval amplitude vs. the
beat number in
the reconstructed, filtered time series. This standard deviation, here denoted
 by $\sigma_{filter}$, can be used to
obtain a separation of sick/healthy subjects.

In fig. \ref{fig1} we display the RR intervals vs. the beat number of a normal
subject. The wavelet technique cleans the highest and lowest frequencies from
the overall picture. The highest frequencies contain noise and the lowest
frequencies
contain mainly external influences on the HR pattern like movement and
slower trends in HR level,
which are not necessarily reflective of autonomic nervous activity. After the
removal of these frequencies one is left with the characteristic frequencies of
the heart.

\begin{figure}[thb]
\psfig{figure=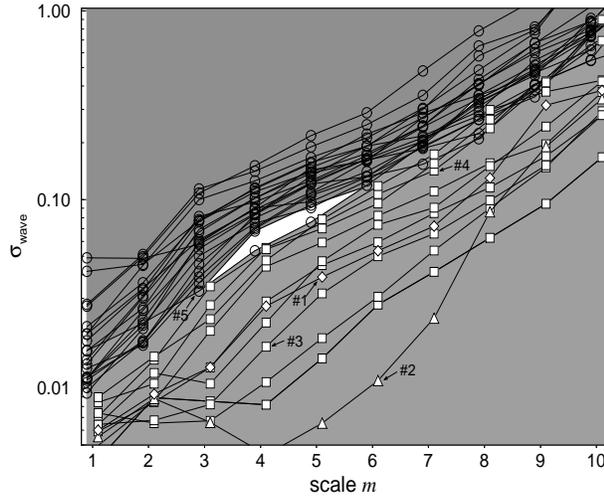,height=7cm,width=10cm,angle=-90}
\caption[]{\label{fig2}
Daubechies 10-tap wavelet. $\sigma_{wave}$, the standard deviation, is plotted 
as a
function of the scale $m$, $1\le m\le 10$. The corresponding window size is 
$2^m$.
The empty symbols indicate the healthy subjects, the opaque symbols indicate
patients. The circles designate normal subjects, the squares - diabetic 
patients, diamond - patient at risk with heart infarct and triangle -
a heart transplanted patient. 
}
\end{figure}

Fig. \ref{fig2} shows the standard deviation $\sigma_{wave}$ for a Daubechies
10-tap wavelet as
a function of the scale number m. The almost total separation between sick and
healthy subjects is obvious.

Patient \#1, falling into the range of sick
patients, has a very low HRV both on a 24-hour scale and short term.
The patient is a survivor of a heart infarct and is at high risk of
sudden cardiac death.

Patient \#2 has the lowest $\sigma_{wave}$ values in the range $4\le m\le6$.
He has undergone a heart transplant; the nerves to the
heart have been disconnected and there is almost no HRV.

Patient \#3 is a diabetic patient, who is classified by the wavelet technique
as a high
risk patient. Diabetic patients with abnormal cardiac autonomic function
have an adverse prognosis and increased risk of heart disease.

Patient \#4, also a diabetic, seems to be less at risk. His $\sigma_{wave}$
 is near the transition between healthy and sick subjects.

The method used in ref. \cite{Thurner98} fails for subject \#5, who appears in
the risk group,
although he had no evidence of diabetes or heart disease.
\begin{figure}[thb]
\psfig{figure=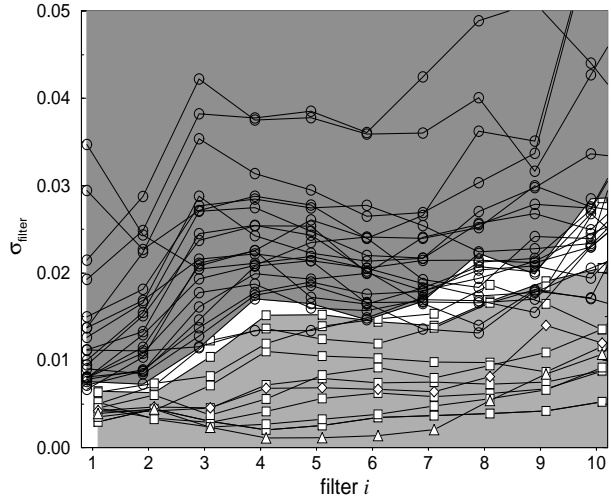,height=7cm,width=10cm,angle=-90}
\caption[]{\label{fig3}
Daubechies 10-tap wavelet filtered inverse transform. The symbols are as
in fig. 2.
}
\end{figure}

In fig. \ref{fig3} the standard deviation of the amplitude of the
reconstructed time
series has been calculated for $1 \le m \le 10$. Again, a total
separation between sick and healthy subjects is apparent. The fact that the
$\sigma_{filter}$ remain almost constant for scales between 4 and 6 for each
individual hints to the possibility that the corresponding frequencies are
characteristic of those at which the autonomic nervous system works.
\begin{figure}[thb]
\psfig{figure=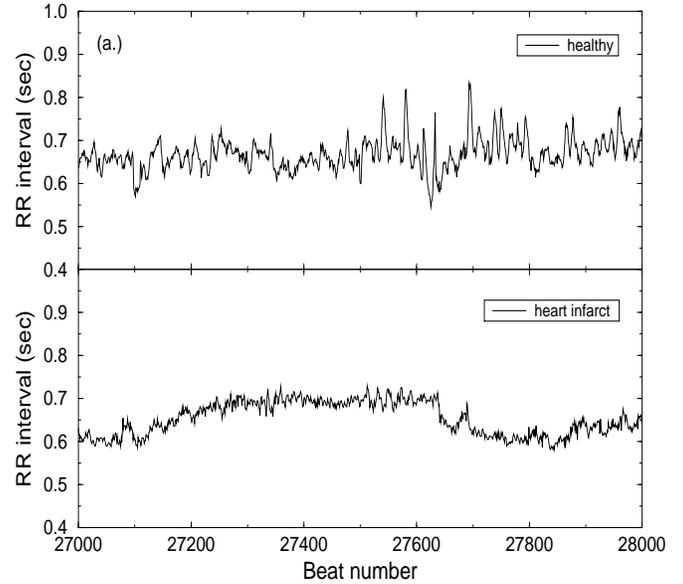,height=8.5cm,width=10cm,angle=-90}
\psfig{figure=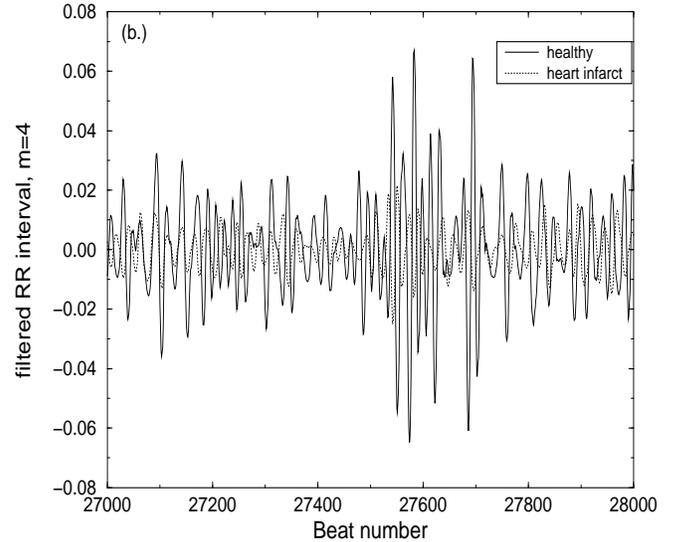,height=8.5cm,width=10cm,angle=-90}
\caption[]{\label{fig4}
(a) Typical time series segments for a sick and a normal individual.
(b) Typical reconstructed, filtered time series for the above individuals.
The segments shown are the same as in (a). The filter is created by the
inverse transform of coefficients with scale $m=4$.
}
\end{figure}

Fig. \ref{fig4}a shows a typical RR interval time series for a healthy and a
sick
subject, whereas fig. \ref{fig4}b shows the reconstructed time series ($m=4$).
One notices
that the difference in amplitudes for healthy/sick subjects is much more
pronounced in the latter  time series.
\begin{figure}[thb]
\psfig{figure=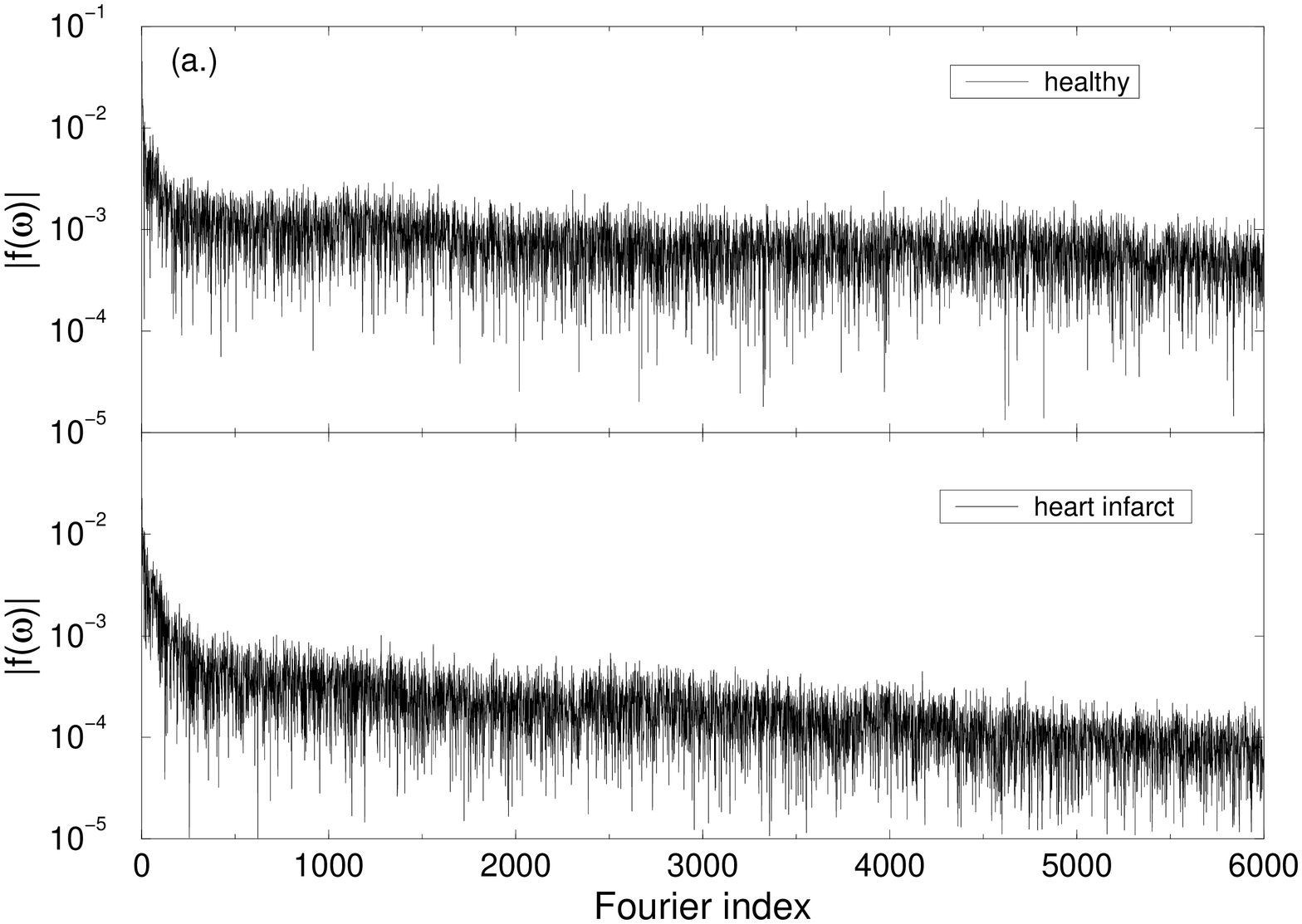,height=8.5cm,width=9cm,angle=-90}
\psfig{figure=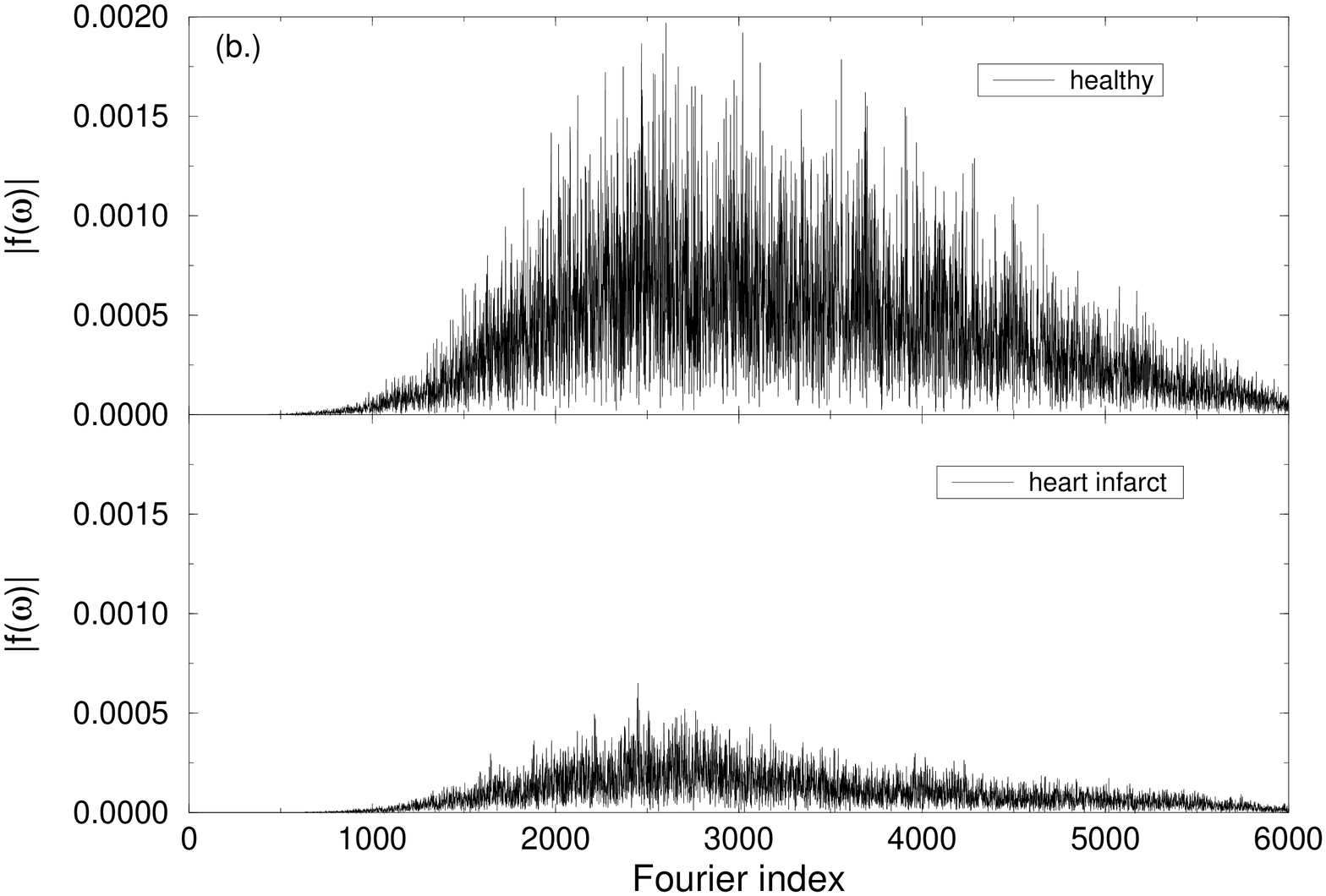,height=8.5cm,width=9cm,angle=-90}
\caption[]{\label{fig5}
(a) and (b). The Fourier transforms of the above (fig. \ref{fig4}). An index
of 1000 represents a frequency of 0.02 Hz.
}
\end{figure}

Figs. \ref{fig5}a and \ref{fig5}b show the Fourier transforms for the time
series displayed in
figs. \ref{fig4}a and \ref{fig4}b, respectively. These power spectra appear
similar, however differ in their respective order of magnitude.
Clearly, the reconstructed filtered time series are distinct by the
amplitude as well as the broadness of their Fourier transforms.

\begin{figure}[thb]
\psfig{figure=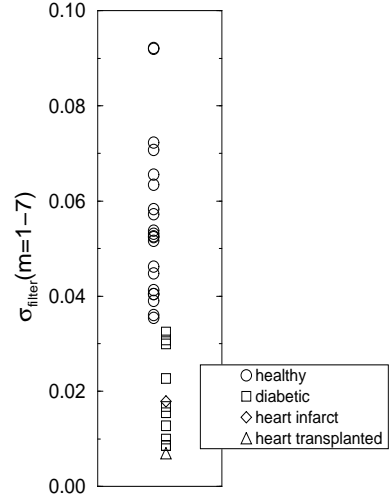,height=8.5cm,width=11cm,angle=-90}
\caption[]{\label{fig6}
Daubechies 10-tap wavelet filtered inverse transform. The symbols are as in
figs \ref{fig2}. The filter is created by the inverse transform of coefficients
with $1\le m \le 7$.
}
\end{figure}

In fig. \ref{fig6} we have obtained a complete separation between the sick and
healthy
subjects by application of a filter which is created by retaining wavelet
coefficients with scales $1 \le m \le 6$. This filter was motivated by the
observation
that a separation is evident for these scales (see figs. \ref{fig2} and
\ref{fig3}). One observes
that the healthy subject \#5, who failed the  wavelet transform diagnostics of
ref. \cite{Thurner98}
 (fig. \ref{fig2}), is now properly classified as not being at risk.

\section{Conclusion}
Our study supports the conjecture of ref. \cite{Thurner98} that healthy
subjects exhibit
greater fluctuations (larger $\sigma_{wave}$ values) than patients.
This difference in fluctuations become most evident on the scale 4 to
5 (corresponding to windows of 16 and 32 heartbeats), but in our study it is
apparent at all scales from 1 to 7 (windows of 2 to 128 heartbeats).

The most distinct difference between sick and healthy individuals appears in
the amplitude changes in the 'reconstructed' time series, where the windows of
16, 32 and 64 heartbeats contribute in a similar way. Letting the window be as
small as $2^4$ heartbeats is enough to allow the healthy group to show
substantial
variation in the size of RR intervals implying a large $\sigma$ value, but is
at the same time too small a window to let the sick cardiac autonomic nervous
system introduce significant
variations in the length of the RR intervals and hence allows it only to reach
a $\sigma$ value essentially smaller than the healthy heart.

The final conclusion of this study is that in order to obtain a complete
separation between healthy subjects and patients one has to consider a
range of scales (as shown in fig. \ref{fig6}) instead of only one scale
(as in figs. \ref{fig2} and \ref{fig3}). This implies that,
$\sigma_{filter}$ as in fig. \ref{fig6} can be used as a
diagnostic indicator, with a delimiting value of $\sim 0.035$ (for the
above mentioned filter), below
which the persons have abnormal cardiac autonomic function and will be at
risk.

\section{Acknowledgments}
M.L. and K.S. are grateful to the Danish-Israel Study Fund in memory of
Josef \& Regine Nachemsohn. Y.A. acknowledges support from the Yad
Jaffah Foundation.


\begin{references}

\bibitem{Moelgaard95} H. Moelgaard, {\it 24-hour Heart Rate Variability.
Methodology and Clinical Aspects}. Doctoral Thesis, University of Aarhus,
(1995).

\bibitem{Furlan90} R. Furlan, S. Guzzetti, W. Crivellaro {\it et al},
Circulation {\bf 81}, 537 (1990).

\bibitem{Akselrod81} S. Akselrod, D. Gordon, F. A. Ubel {\it et al},
Science {\bf 213}, 220 (1981).

\bibitem{Pomeranz85} B. Pomeranz, R.J.B. Macaulay, M.A. Caudill {\it et al},
Am. J. Physiol. {\bf 248}, 151 (1985).

\bibitem{Moelgaard94} H. Moelgaard, P.D. Christensen, H. Hermansen
{\it et al}, Diabetologia {\bf 37}, 788 (1994).

\bibitem{Bigger91} J.T. Bigger, J.L. Fleiss, L.M. Rolnitzky {\it et al},
JACC {\bf 18}, 1643 (1991).

\bibitem{Task96} {\it Task force of ESC and NASPE}, Eur. Heart J., 354 (1996).

\bibitem{Daubechies92} I. Daubechies, Ten Lectures on Wavelets
(Society for Industrial and Applied Mathematics, Philadelphia, PA 1992)

\bibitem{Strang96} G. Strang and T. Nguyen, {\it Wavelets and Filter Banks},
(Wellesley-Cambridge Press, Wellesley, 1996)

\bibitem{NR} W. H. Press, S. A. Teukolsky, W. T. Vetterling and B. P.
Flannery, {\it Numerical Recipes in C}, 2nd Ed., Cambridge University,
Cambridge 1995.

\bibitem{Aldoubri96} A. Aldoubri and M. Unser, eds., Wavelets in Medicine and
Biology (CRC Press, Boca Raton, FL, 1996)

\bibitem{Akay97} M. Akay, ed, Time Frequency and Wavelets in Biomedical Signal
Processing (IEEE Press, Piscataway, NJ, 1997)

\bibitem{Ivanov96} P.C. Ivanov, M.G. Rosenblum, C.-K. Peng, J. Mietus,
S. Havlin, H.E. Stanley, and A.L. Goldberger, Nature {\bf 383}, 323 (1996)

\bibitem{Thurner98} S. Thurner, M.C. Feuerstein and M.C. Teich,
Phys. Rev. Lett. {\bf 80}, 1544 (1998).

\bibitem{Peng95} C.K. Peng, S. Havlin, H.E. Stanley and A.L. Goldberger,
Chaos {\bf 5}, 82-87, (1995)

\end{references}
\end{document}